**Monte Carlo Characterization of the Cosmic Ray Muon Flux in Subsurface Geological Repositories**

Harish Gadey[1, *], Stylianos Chatzidakis[2], and Abi T. Farsoni[1]

[1]School of Nuclear Science and Engineering, Oregon State University, Corvallis, Oregon, USA 97331

[2]Oak Ridge National Laboratory, Oak Ridge, Tennessee, USA 37831

[*]**Contact Info:**

Tel: (541) 602-2370, Email: gadeyh@oregonstate.edu

**Abstract**

Recent challenges in monitoring subsurface geological repositories call for new, innovative concepts that are facility independent, cost-effective, passive, and reliable. Inspection and verification of future disposal facilities will exert significant pressure on the limited safeguards resources. Compared to aboveground facilities, subsurface geological repositories cannot be directly monitored. Once nuclear material is in place in these facilities, reverifying the inventory may no longer be feasible if continuity of knowledge is lost or updated safeguards information on the contents (or lack thereof) becomes available to inspectors. Using cosmic ray muons presents several potential advantages over conventional photon/neutron signatures, and their use in safeguards applications has only recently received attention. However, there have been limited efforts to explore the integration of cosmic ray muons into repository safeguards and study potential gains, risks, and costs. This paper presents a Monte Carlo-based methodology to characterize the cosmic ray muon flux, including muon angular and energy differential distributions at depths representative of geological repositories. This work discusses the feasibility of muon monitoring for detecting spent nuclear fuel disposal cask movement or for unauthorized excavation and rock removal. The objective is to develop useful parametrizations to provide a convenient tool for detector-specific and safeguards applications at any geological repository site. It is expected these results will provide a better understanding of how muons can be integrated into an existing geological repository safeguards framework.

Notice:  This manuscript has been authored by UT-Battelle, LLC, under contract DE-AC05-00OR22725 with the US Department of Energy (DOE). The US government retains and the publisher, by accepting the article for publication, acknowledges that the US government retains a nonexclusive, paid-up, irrevocable, worldwide license to publish or reproduce the published form of this manuscript, or allow others to do so, for US government purposes. DOE will provide public access to these results of federally sponsored research in accordance with the DOE Public Access Plan (http://energy.gov/downloads/doe-public-access-plan).

# I. INTRODUCTION

Cosmic ray muon monitoring is a promising next-generation technology for nondestructive assay. The natural generation of relativistic muons, produced in the atmosphere by cosmic rays, and their ability to penetrate through dense materials indicate that the technique could be an excellent candidate for efficient, inexpensive, remote safeguarding of subsurface geological repositories and disposal casks. Many monitoring and imaging applications based on cosmic ray muons have been investigated over the years, with emphasis on archaeology, volcano imaging, and material identification. E. P. George [1], who measured cosmic ray muon flux attenuation to infer rock depth covering underground tunnels, and L. Alvarez et al. [2], who searched for hidden chambers within the Egyptian pyramids, both contributed pioneering work in this area. Recent efforts led to the development of imaging algorithms for scanning cargo, locating molten fuel, and analyzing spent fuel [3–7]. Los Alamos National Laboratory (LANL) and the National Institute for Nuclear Physics (INFN) demonstrated the use of muons and multiple Coulomb scattering to detect hidden high Z materials among a large volume of low-Z materials and developed muon detectors suitable for surveillance of cross-border transport containers [8, 9]. Potential muon monitoring applications could be used to verify international treaty declarations to ensure that nuclear material diversion has not occurred, including removal of spent nuclear fuel from dry casks or disposal canisters.

As of 2019, several countries have confirmed plans for final disposal of spent nuclear fuel and other nuclear materials in subsurface geological repositories [10–13]. Unlike above-ground facilities, subsurface geological repositories cannot be directly monitored. Once nuclear material is in place, reverifying the inventory may no longer be feasible if continuity of knowledge is lost or updated safeguards information on the contents (or lack thereof) becomes available to inspectors. Consequently, safeguards of subsurface geological repositories will require nonconventional technologies capable of near real-time, unattended, very long-term monitoring. The extreme attenuation of neutrons/photons at these depths renders common monitoring mechanisms ineffective. This problem can be partially addressed using cosmic-ray muons that are not only highly penetrating in nature but are also continuously generated in the upper atmosphere. Muon detection technology has already been demonstrated in various real-world applications [3–9], and it could offer a solution to this complex problem by integrating the recent advances in cosmic ray muon detection and monitoring within the conventional repository safeguards framework. Early consideration and integration of cosmic ray muons within the safeguards' framework will ultimately facilitate an effective, cost-efficient solution.

The characterization of cosmic ray muon flux as a function of depth is critical to the design of underground physics laboratories and has been extensively studied at several underground facilities globally. Measurements of the muon flux per unit solid angle as a function of depth have been used for background estimations at underground facilities for neutrino detection [14], dark matter studies [15, 16], and other rare-events experiments [17]. Recently, Monte Carlo simulations and curve-fitting tools have been used to investigate and predict vertical muon intensity and angular distributions at large depths [18–20]. However, available measurements and simulations typically focus on depths larger than 1,000-m water equivalent (m.w.e.) (1 km.w.e. = $10^5$ g/cm$^2$), and additional studies are needed to fully characterize cosmic ray muon flux at geo-repository depths (500–1500 m.w.e.). This paper presents a Geant4 (for **ge**ometry **an**d **t**racking)-based methodology to characterize the cosmic ray muon flux, including angular and energy differential distributions, at depths representative of geological repositories [21-24]. The objective is to develop useful parametrizations to provide a convenient tool for detector-specific and nonproliferation applications at any geological repository site. The results are anticipated to provide a better understanding of how muons can be integrated into an existing geological repository safeguards framework.

## II. BACKGROUND

The processes occurring in the atmosphere lead to the creation of a cascade of secondary rays and relativistic particles. Among these particles, pions and kaons decay to muons and cause a considerable muon flux that reaches the sea level. Cosmic ray muons are unstable particles of two charge types, weighing approximately 200 times the mass of an electron and they rain down on earth's surface [25–28]. High-energy muons show no azimuthal dependence. However, since muons with a higher zenith angle lose more energy while propagating through matter, there is a significant zenith angle dependence, resulting in a smaller intensity of muons with a large zenith angle compared to vertical muons (low zenith angle). Knowledge about the incident muon flux is critically important in determining the underground intensity and any bias in the incident flux will lead to a bias in the estimated underground intensity.

### A. Main muon flux characteristics

#### 1. Surface level

Muons reach the earth's surface with an average energy of 3–4 GeV at an approximate rate of 10,000 particles $m^{-2}$ $min^{-1}$. The muon spectrum at the surface has been experimentally measured and shown to vary significantly with energy and zenith angle. A large pool of experimental data exists to determine the surface flux for energies up to 1 TeV and zenith angles between 0 and $\pi/2$. Muon spectra at sea level for various zenith angles are shown in Fig. 1. Above 100 GeV, the spectrum follows a power-law profile that is practically independent of the zenith angle. Below 100 GeV, the spectrum slope changes, and the zenith angle dependence increases. The overall zenith angular distribution is approximately proportional to that of cosine squared. The influence of the zenith angle is small up to 30° but becomes significant for larger angles. The muon spectrum is also influenced by geomagnetic effects, which limit the primary proton flux, and the solar cycle, which modulates the primary proton flux [29–32]. These dependencies become significant for muon energies less than 10 GeV. With an increase in altitude, these variations become less significant [33]. Despite a large amount of data available, significant differences still exist among models published by several authors.

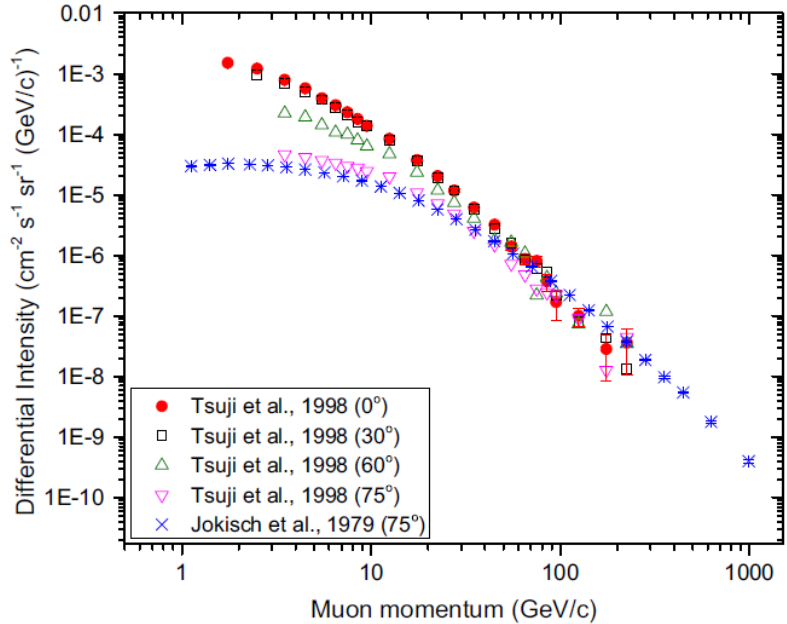

FIG. 1. Muon differential intensity at sea level for different zenith angles from Tsuji et al. and Jokisch et al. [34, 35].

Muons can penetrate through several tens to hundreds of meters of rock. Muons that penetrate deep underground typically have energies higher than 100 GeV at sea level. Eq. (1) shows a widely used approximate formula for sea level muon flux with energies higher than 100 GeV/cosθ and zenith angles θ < 70° [36].

$$\frac{dN\mu}{dE\mu\, d\Omega} = \frac{0.14 E\mu^{-2.7}}{cm^2\, s\, sr\, GeV} \cdot \left\{ \frac{1}{1+\frac{1.1\, E\mu\, \cos\theta}{115\, GeV}} + \frac{0.054}{1+\frac{1.1\, E\mu\, \cos\theta}{850\, GeV}} \right\}, \quad (1)$$

where the two terms give the contributions from pions and kaons, $E\mu$ is the muon energy in GeV, and $\theta$ is the zenith angle. A major limitation of Eq. (1) is that it strongly overestimates the muon flux for energies less than 100 GeV. To overcome this limitation, the analytical model proposed by Smith and Duller [37] can be used.

### *2. Deep underground*

Cosmic ray muons propagating through matter lose energy via inelastic collisions with electrons and radiative processes (bremsstrahlung, nuclear, pair production) and suffer deflection from nuclei due to multiple Coulomb scattering. For an underground site, the differential energy spectrum at depth *h* is given in Eq. (2) [36]:

$$\frac{dN\mu(h)}{dE\mu\, d\Omega} = \frac{dN\mu}{dE_{\mu,0} d\Omega} e^{bh}, \quad (2)$$

where $E_{\mu,0}$ is the muon energy at sea level and *b* is the energy loss fraction from radiative processes (see Table I). Eq. (3) shows the differential muon intensity as a function of depth and zenith angle [19]:

$$I(h,\theta) = I(0^0, h)\cos^n(\theta), \quad (3)$$

where *I* is the through-going muon intensity, *I(0°,h)* is the vertical intensity, *n* is a depth-dependent quantity, and *θ* is the zenith angle. Total muon flux underground is shown in Eq. 4 [18]:

$$I_{tot} = \int \sin(\theta)\, d\theta \int d\phi\, I(h(\theta,\phi))G(h,\theta), \quad (4)$$

where $I_{tot}$ is the total muon flux, *G(h,θ)* equals *sec(θ)*, and *h(θ, φ)* provides the topology of the geometry above. The statistical energy average loss of muons traversing X amount of matter with energies considerably higher than the Bethe-Bloch minimum is shown in Eq. (5) [36]:

$$-\frac{dE\mu}{dX} = a(E\mu) + b(E\mu)\, E\mu, \quad (5)$$

where *a* represents the ionization loss and *b* represents the energy loss fraction from radiative processes. The main parameters influencing *a* and *b* are the average atomic over mass number <Z>/<A> and the bulk density of the material. The average values of *a* and *b* along with the range of muons for a given energy in standard rock are shown in Table I [36].

TABLE I. Range, ionization loss and energy loss fraction of muons in standard rock for different energies.

| Muon energy at surface level (GeV) | Range (km.w.e.) | *a* (MeV cm²/g) | *b* (brems) ($10^{-6}$ cm²/g) | *b* (pair-prod) ($10^{-6}$ cm²/g) | *b* (nuc) ($10^{-6}$ cm²/g) | *Total b* ($10^{-6}$ cm²/g) |
|---|---|---|---|---|---|---|
| 10 | 0.05 | 2.17 | 0.70 | 0.70 | 0.50 | 1.90 |
| 100 | 0.41 | 2.44 | 1.10 | 1.53 | 0.41 | 3.04 |
| 1,000 | 2.45 | 2.68 | 1.44 | 2.07 | 0.41 | 3.92 |

In this work, muons with energies in the range of 25–1000 GeV were generated in a simulation study to accurately estimate the muon background at subsurface geological repositories. If the energy dependency of parameters *a* and *b* are neglected, then Eq. (5) can be easily solved, and the minimum initial energy necessary for a muon to cross a given amount of matter can be used to compute the integrated flux underground. However, the accurate determination of these functions and the muon intensity requires modeling of cross-sections and particle interactions feasible only through Monte Carlo simulations.

## B. Main characteristics of geological repositories

Table II shows publicly available rock compositions and the average depth of subsurface geological repositories at various locations. Rock density at these locations varies between 1.87–2.75 g/cm$^3$, and depth ranges from 300 m to slightly over 400 m. The approximate average atomic weights and numbers are calculated based on known local rock composition. All repositories share a similar $<Z>/<A>$ ratio, and the only varying quantity is density. This important observation allows us to use *a* and *b* parameters and muon flux computed for standard rock and the results may safely be used to determine the muon flux across all repositories correcting only for differences in bulk density.

Besides the ones mentioned in the table, repositories are being planned to be constructed in France, Canada, and Switzerland, among other countries, all of which are in their early siting or feasibility study phase.

Table II. Average properties of various geological repositories.

| Site | Type | Effective atomic mass, $<A>$ | Effective atomic number, $<Z>$ | $<Z>/<A>$ | Average density (g/cm$^3$) | Depth (m) | Meter water equivalent (m.w.e.) |
|---|---|---|---|---|---|---|---|
| Finland | Granite[a] [38] | 26.69 | 13.04 [39] | 0.488 | 2.75 [38] | 420 | 1,155 |
| Sweden | Granite[a] [40] | 26.69 | 13.04 [39] | 0.488 | 2.70 [40] | 400 | 1,080 |
| USA | Clay/Shale[b] [41] | 20.56 | 10.38 | 0.504 | 1.87 [41] | 300 | 561 |
| WIPP [42] | Salt[c] | 30.0 | 14.64 | 0.488 | 2.3 | 689 | 1,585 |
| This study | Standard rock[d] | 22 | 11 | 0.5 | 2.65 | 300–500 | 795–1325 |

[a] Granite composition: $SiO_2$, $Al_2O_3$, $K_2O$, $Na_2O$, CaO
[b] Clay composition: $Al_4Si_6O_{15}(OH)_2$
[c] Salt composition: NaCl
[d] Standard rock composition: $CaCO_3$, $MgCO_3$

## III. GEOLOGICAL REPOSITORY MODELING METHODOLOGY

Geant4 was used for simulating muon transport through rock. Geant4 is a Monte Carlo code designed and developed by the high-energy physics community for tracking subatomic particles and their interactions with matter. For repository modeling, standard rock with a density of 2.65 g/cm$^3$ and a mix of $CaCO_3$ and $MgCO_3$ with mass fractions of 52% oxygen, 27% calcium, 12% carbon, and 9% magnesium were used [43]. A cuboid box with material properties of standard rock and variable thickness to represent depth variation was employed to mimic the volume of the repository. A detector was simulated at various repository depths to record the counts and obtain angular and energy information. Geant4 simulations required knowledge of the muon spectrum and angular distribution at sea level, as well as the ability to repeatedly generate random samples from these distributions. Geant4 does not provide a built-in library for muon energy and angular distribution. These muon distributions were obtained using the MUFFSgenMC, an open-source muon event generation code available on the MathWorks website [44]. Data on the number of muon interactions, muon histories simulated, muon source-detector solid angles, energy and angular correction factors, and area of the detection surface were used to obtain the value of the muon flux at each depth. High-performance computing (HPC) resources were used to speed up the simulations. Using 32 CPUs (1 node), one could simulate close to 3.3 million muons per hour. Using 8 nodes, enabled running, on average, 60 times faster compared to a dual-core, 4-thread processor.

## A. Muon generation and solid angle

The solid angle of the detector with respect to the muon generation source plane was maintained constant (60°) at variable depths as suggested by Arslan and Bektasoglu [19]. To keep the same solid angle in all simulations independent of depth, the muon generation cone radius is calculated according to Eq. (6):

$$Muon\ generation\ cone\ radius\ (m) = tan(60°) * repository\ depth\ (m) \quad (6)$$

The energy of the muons simulated in this work was in the range of 25–1,000 GeV. Since the solid angle of the detector was kept at 60°($\theta_{SA}$), muons with a zenith angle range of 0–60° were generated in all simulations independent of depth. A flowchart is presented in Fig. 2 to illustrate the overall methodology. Fig. 3(a) shows a sketch of the repository with the muon detector placement and Fig. 3(b) shows a Geant4 visualization.

## B. Post-processing

After the geometry was modeled and macro files were updated based on the energy and angular distribution of interest, the Geant4 executable was run until a predetermined number of events was simulated. The output was then parsed using a MATLAB script to extract energy and angular data. Energy and zenith angle correction factors were applied while calculating the flux at a given depth, and muon energies (1–25 GeV) and zenith angles (60°–89°) that were not simulated were considered. These correction factors were calculated based on the initial energy and angular spectra obtained from the Smith and Duller analytical solution. To speed up the simulation, all secondary particles generated during the simulation were terminated.

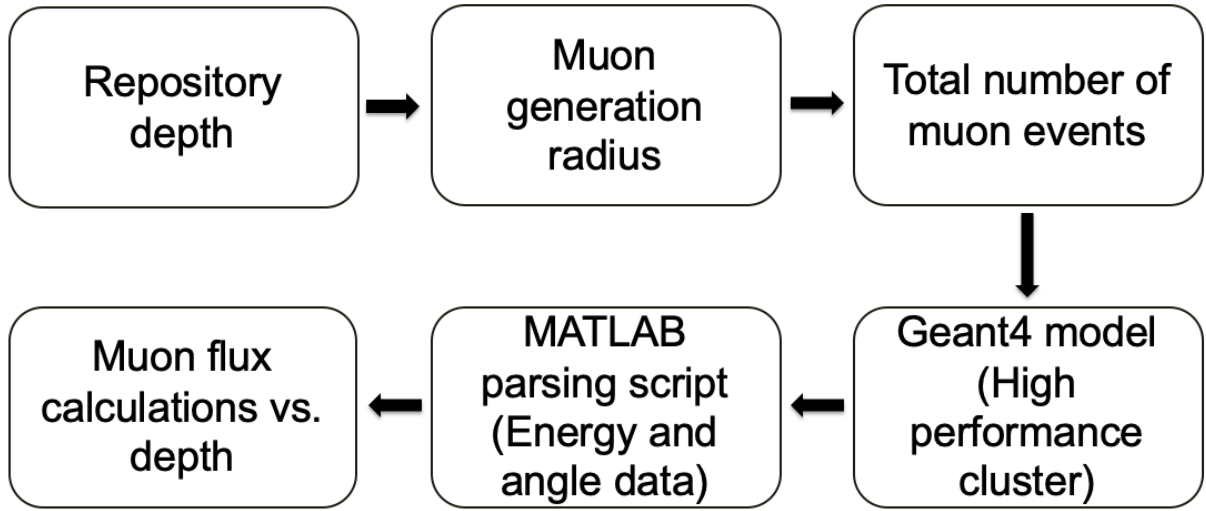

FIG. 2. Flowchart depicting the methodology for calculating the muon flux and extracting angular data.

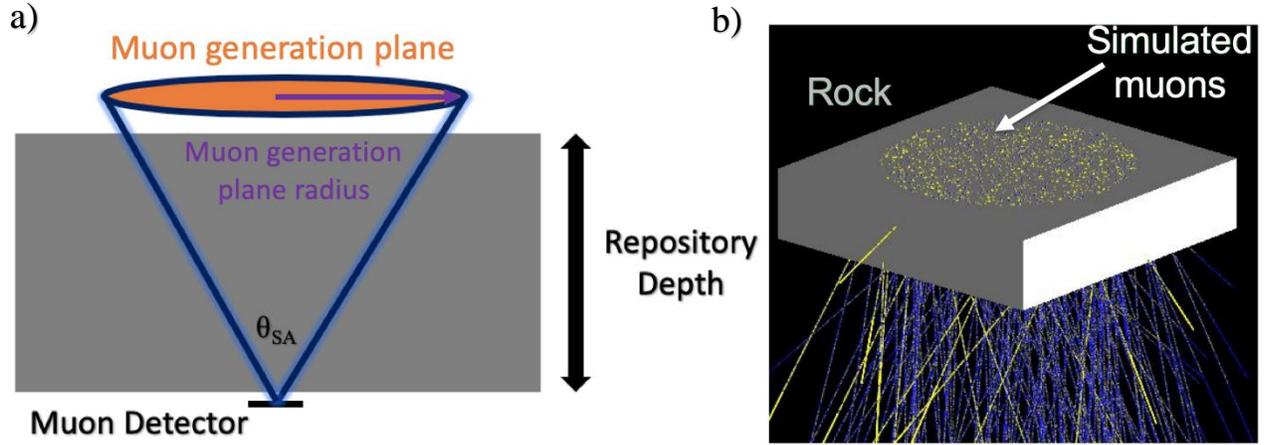

FIG. 3. (a) Sketch of the repository depicting the placement of the muon detector and the muon generation plane. (b) Geant4 graphical user interface model of the repository showing the particle generation surface and the muon trajectories.

## IV. MODEL BENCHMARKING

Prior to performing muon-repository simulations using Geant4, the model was first benchmarked and simulation results were compared with publicly available experimental measurements. Fig. 4(a) shows simulated data at 100, 400, 700, and 1,325 m.w.e. along with experimental measurements. A correlation that models the muon flux from 100 m.w.e. to about 1,300 m.w.e. was developed based on Geant4 simulations results (Eq. 7). The correlation was compared with experimental results at each respective depth [Fig. 4(b)]. It is noted that the muon energy for the first three Geant4 data points was limited to 1 TeV due to validation constraints of the muon event generator, although in reality, the energy of muons can extend well beyond 1 TeV. This limited energy range resulted in large deviations from experimental measurements at larger depths, i.e., 1,325 m.w.e. Simulations were then performed with muon energies up to 60 TeV; these simulations overestimated the muon flux by almost 700%. This observation revealed the strong dependence of the maximum muon energy on the estimated underground cosmic ray muon flux at large depths. This effect is not pronounced at shallow depths because the energy spectra at these levels are dominated by low-energy muons. A second study was subsequently carried out using muons with a maximum energy of 20 TeV at a depth of 1,325 m.w.e. that resulted in an estimated muon flux within 15.3% of the experimental results. Therefore, the extended energy range was used for larger depths.

$$I_u(h) = Ae^{bh} + Ce^{dh}, \qquad (7)$$

Where, A is $(0.0003783 \pm 0.0000002)$ cm$^{-2}$ sec$^{-1}$ sr$^{-1}$, b is $(-0.009737 \pm 0.000025)$ m.w.e.$^{-1}$, C is $(1.829 \pm 0.03) \times 10^{-6}$ (cm$^{-2}$ sec$^{-1}$ sr$^{-1}$), d is $(-0.00113 \pm 0.0023)$ m.w.e.$^{-1}$ and, h is depth in m.w.e.

It is noted that in Fig. 4(a), the average altitude where the seven experiments were performed was approximately 800 m. The Geant4 results were adjusted to account for this altitude increase since the muon event generator provides sea-level muon distributions. Eq. (8) presents the muon count rate (cpm) as a function of altitude [45]:

$$Muon\ count\ rate(x) = ae^{bx}, \qquad (8)$$

where a is 95.641, b is 0.3211, x represents the altitude in kilometers, and *muon count rate(x)* represents the number of muons recorded per minute at *x* kilometers above sea level. The increase in muon counts can be used to adjust the simulation results to account for altitude. Another correlation suggested that for every 1,000-m increase in altitude from mean sea level, the muon flux increases by about 10% [46]. Both methods yielded approximately the same result. The developed correlation captures correctly the muon behavior as it changes to a less steep profile at larger depths but consistently underestimates both the measured and simulated muon intensity by 10-60%. It is worth noting that individual simulations tend to have a better representation of the muon intensity and are closer to the measurements than the developed correlation. Overall, the developed Geant4 model appears to be in good agreement with measurements at various depths given the lack of detailed information about the experimental conditions, e.g., exact measurement location, overburden shape and material, etc., that unavoidably lead to a simplified model.

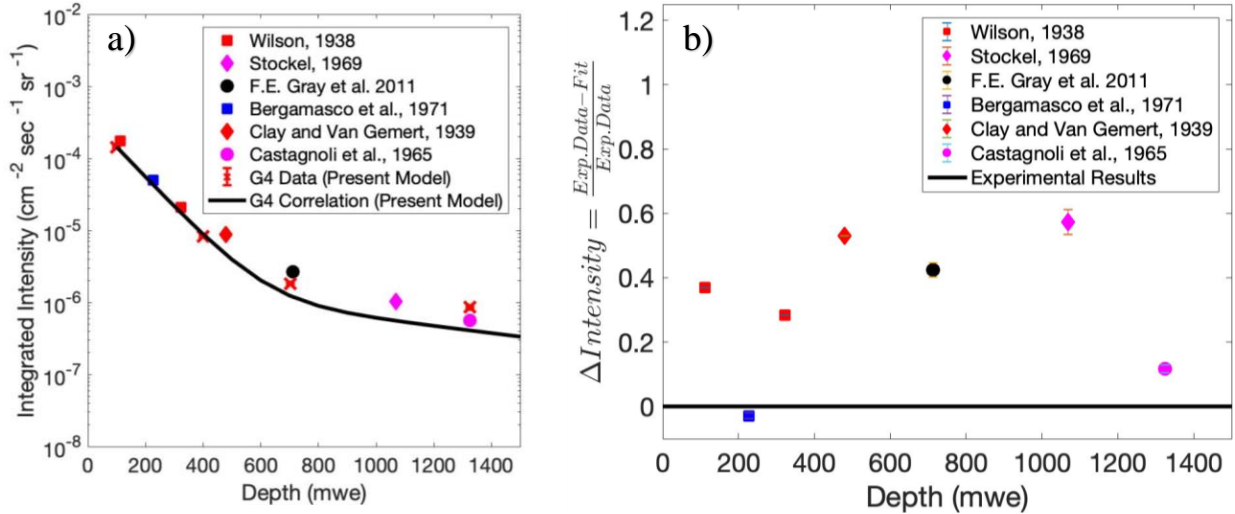

FIG. 4. (a) Geant4 data and correlation vs. experimental flux data. (b) Altitude adjusted relative experimental flux variation from the developed correlation. Note: experimental error bars where available are included in both figures but are smaller than the plot symbols; simulation error bars represent statistical uncertainty at one standard deviation.

## V.   RESULTS AND DISCUSSION

### A.  Muon intensity versus depth

This section presents simulation results at variable repository depths. This work simulated muon flux at 5 depths between 300–500 meters of rock and parametrized correlations for muon flux are developed in this range. The muon zenith angle was also studied at various depths to understand the change in angular distribution shape with respect to sea level. Table III presents the muon flux obtained from Geant4 simulations at various depths, with associated errors.

TABLE III. The muon flux and associated errors at various repository depths.

| Depth-standard rock (m) | Depth (m.w.e.) | Muon flux (cm$^{-2}$ s$^{-1}$ sr$^{-1}$) | Error (cm$^{-2}$ s$^{-1}$ sr$^{-1}$) | Ratio (Surface/underground)[a] |
|---|---|---|---|---|
| 300 | 795 | 9.82E-07 | 0.23E-07 | 2,701 |
| 350 | 927.5 | 6.17E-07 | 0.25E-07 | 4,302 |
| 400 | 1,060 | 3.85E-07 | 0.60E-07 | 6,886 |
| 450 | 1,192.5 | 2.63E-07 | 0.13E-07 | 10,072 |
| 500 | 1,325 | 1.63E-07 | 0.10E-07 | 16,235 |

[a] Solid angle of $2\pi$ is used for surface flux calculations.

A correlation was developed in this repository range that describes the flux profile with an R-squared value of 0.999:

$$I_u(h) = Ae^{bh}, \quad (9)$$

where $A$ is $(1.464 \pm 0.414) \times 10^{-5}$ (cm$^{-2}$ sec$^{-1}$ sr$^{-1}$), $b$ is $-(0.003405 \pm 0.000334) \times 10^{-4}$ (m.w.e.$^{-1}$), and $h$ is depth in m.w.e. Fig. 5 shows the obtained results and a comparison with a correlation developed by Arslan and Bektasoglu [19]:

$$I_\mu(h) = A\left(\frac{h0}{h}\right)^2 e^{-h/h0}, \quad (10)$$

where $A$ is $(0.89 \pm 0.07) \times 10^{-6}$ cm$^{-2}$ sec$^{-1}$ sr$^{-1}$, $h_0$ is $1{,}307 \pm 3$ m.w.e., and $h$ is the depth in m.w.e. This equation was developed to model cosmic ray muon flux for underground laboratories at depths greater than 4,000 m.w.e.

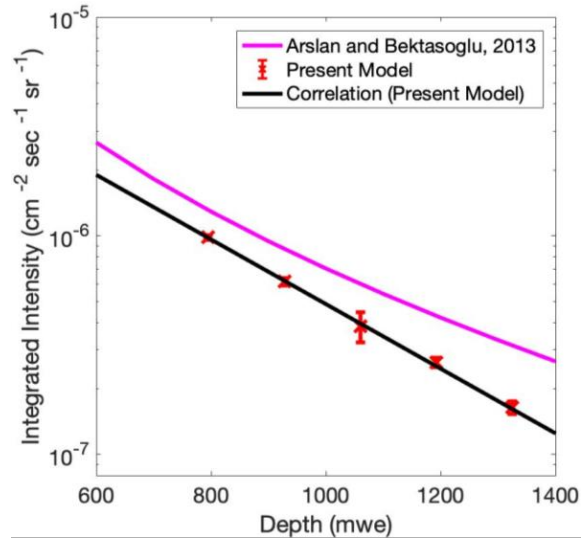

FIG. 5. Developed correlation for repository depths.

It is observed that Eq. 10 consistently overestimates the muon flux and does not correctly capture the shape at the depths of interest. The considerable difference in shape and magnitude (between 40 and 90 percent) observed between the Arslan and Bektasoglu correlation and the present model (Fig. 5) demonstrates the need for repository specific correlations.

### B. Angular distribution

Fig. 6(a) shows the calculated relative intensity variation in the zenith angle at various depths. The figure shows that with an increase in depth, there is a rapid reduction in intensity for the same zenith angle. This is because, with an increase in zenith angle, a muon is forced to traverse through a larger amount of rock to reach the detector. Therefore, there is a decrease in muon intensity with an increase in repository depth and this change is more pronounced at higher zenith angles. For example, at 0º zenith angle the variation with depth is negligible, at 40º zenith angle, the muon intensity is reduced approximately 30% at 265 m.w.e. compared to that at sea level. Although at larger depths the variation is less pronounced, the zoomed-in image of the high zenith angle profile in Fig. 6(a) (inset) illustrates that this phenomenon holds true at all depths.

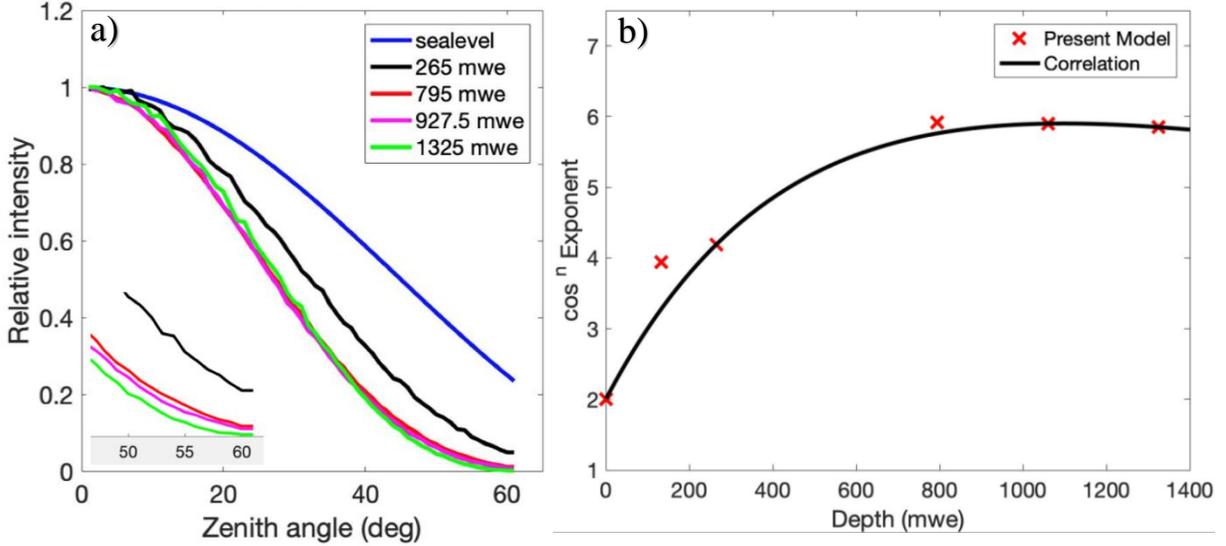

FIG. 6. (a) Relative intensity variation with zenith angle as a function of depth. The insert figure shows the subtle differences at larger depths. (b) cosine exponent curve as a function of depth (m.w.e.).

The cosine exponent of the zenith angle profile as a function of depth was determined using a recursive function that yielded the least deviation between the cosine curve and the zenith angle profile at each depth. An empirical function for the exponent (*n*) was then determined using a curve fitting tool and is presented in Eq. (11).

$$n = ae^{bx} + ce^{dx}, \tag{11}$$

where *a* is 7.953, *b* is -0.000188, *c* is -5.953, *d* is -0.00214, and *x* is depth in m.w.e. The plot of the cosine exponent curve as a function of depth is shown in Fig. 6(b). The exponent increases monotonically from 2 at sea level to 6 at 1400 m.w.e. This result could be used to allow faster simulation of muons by starting the muon at an initial depth using the corresponding cosine exponent at that depth instead of the more common approach of starting the muons at sea level which typically requires propagation through matter to reach the desired depth.

### C. Differential energy distribution and minimum energy

The differential muon energy distribution is plotted at different depths in the zenith angle range of 0–60°, as shown in Fig. 7(a). It is observed that the muon spectrum moves to a higher minimum and average energies at larger depths. Theoretically, the minimum muon energy can be calculated according to Eq. (12) [47]:

$$E_{0,min} = \varepsilon_\mu \big(e^{\beta X} - 1\big), \tag{12}$$

where $E_{0,min}$ is the minimum muon energy, $\varepsilon_\mu$ is the critical energy where ionization energy loss equals radiative energy loss ~ 500 GeV, $\beta$ is the fractional energy loss in radiative processes ($4 \times 10^{-6}$ cm$^2$/g), and *X* is the depth in g/cm$^2$. The normalized initial muon energy distribution obtained from Geant4 simulations as a function of depth is shown in Fig. 7(b). This information could expedite underground simulations significantly by negating all muon energies in a distribution that falls below a certain energy threshold.

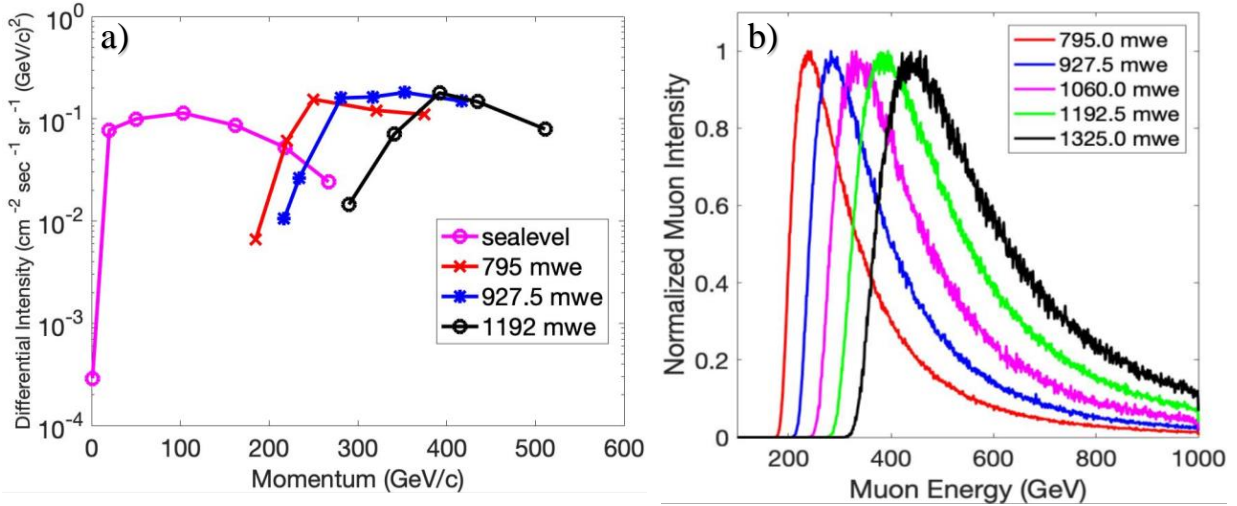

FIG. 7. (a) Differential muon intensity plot as a function of incident muon energy. (b) Normalized initial muon intensity as a function of minimum initial muon energy.

Table IV shows a comparison between theoretical and Geant4-obtained minimum muon energies required to penetrate various depths. Good agreement is observed with a relative difference of less than 10% in all cases.

TABLE IV. A comparison of theoretical and Geant4-obtained minimum muon energies as a function of depth.

| Depth (m.w.e.) | Theoretical (GeV) | Present Model (GeV) |
|---|---|---|
| 795.0 | 187 | 181 |
| 927.5 | 225 | 215 |
| 1,060.0 | 264 | 249 |
| 1,192.5 | 306 | 286 |
| 1,325.0 | 350 | 322 |

## VI. APPLICATION TO MONITORING SUBSURFACE GEOLOGICAL REPOSITORIES

In this section, we present muon flux conditions and possible applications of muon monitoring for subsurface geological repositories. Using the developed muon intensity correlation to model the cosmic ray muon flux at repository depths, values were calculated for repositories in Finland, Sweden, and the United States. Using a solid angle of $2\pi$, the total muon flux at these underground locations was also obtained (Table V). The values were altitude adjusted for the Yucca Mountain Nuclear Waste Repository, which is 2,044 m above sea level, and the Waste Isolation Pilot Plant (WIPP), which is 1,004 m above sea level. The other two sites are situated at sea level, so no altitude adjustments were made.

TABLE V. Estimated vertical muon intensity and total muon flux at four subsurface geological repositories.

| Site | Depth (m.w.e) | Vertical intensity (cm$^{-2}$ s$^{-1}$ sr$^{-1}$) | Total flux (cm$^{-2}$ s$^{-1}$) |
|---|---|---|---|
| Finland | 1,155 | $2.8679 \times 10^{-7}$ | $1.8020 \times 10^{-6}$ |
| Sweden | 1,080 | $3.7023 \times 10^{-7}$ | $2.3262 \times 10^{-6}$ |
| USA | 561 | $4.1783 \times 10^{-6}$ | $2.6253 \times 10^{-5}$ |
| WIPP | 1,585 | $6.6326 \times 10^{-8}$ | $5.7527 \times 10^{-7}$ |

Table V values could be used for feasibility studies in a safeguards framework, e.g., in the calculation of time needed to detect density variations in a repository from high-density material movement. For example, assume a subsurface repository at 400 m where a muon detector with acceptance T= 1000 cm$^2$sr is placed below spent nuclear fuel disposal casks. A density variation of 2000 g/cm$^2$ (assuming average disposal cask density = 10 g/cm$^3$ and diameter 2 m) would result in muon intensity variation of $\Delta I \approx 0.05$ cm$^{-2}$ sr$^{-1}$/day that can be resolved with a measurement duration of a few hours. Small variations of rock density that could indicate post-closure excavations around the repository could also be monitored. A variation in rock depth of 200 g/cm$^2$ (2 m x 1 g/cm$^3$) would give an intensity variation of $\Delta I \approx 0.00098$ cm$^{-2}$ sr$^{-1}$/day. To resolve this density variation, a measurement duration of approximately 30-40 days would be needed, which is well within the timeframe of a repository. Of course, measurement duration can be decreased by using a larger detector or multiple detectors. These simple examples demonstrate the possible use of muon monitoring for detecting disposal cask movement or unauthorized excavation and removal of rock.

## VII. CONCLUSIONS

In this paper, we performed Monte Carlo simulations for the characterization of cosmic ray muon flux as a function of depth at subsurface geological repositories using Geant4. We presented a Geant4-based methodology to characterize the cosmic ray muon flux, including angular and energy differential distributions, at depths representative of geological repositories, and developed parametrizations that can be used to predict muon related quantities at standard rock for depth range 300 – 500 m (795 – 1325 m.w.e.). An important observation that repositories share a similar <Z>/<A> ratio allows to compute the muon flux for standard rock and then converting the result to any repository material correcting only for differences in bulk density.

The Geant4 model was benchmarked against available experiments at various depths and good agreement was observed providing confidence to the correctness of the model to capture the main phenomena. The cosine exponent of the zenith angle profile as a function of depth was determined and it was found that it increases monotonically from 2 at sea level to 6 at 1400 m.w.e. The depth intensity and cosine exponents correlations could be used to simulate underground muons, eliminating the need to transport them through the entire volume of the repository, thereby saving computing resources and efforts. Finally, based on the results obtained, possible applications of muon monitoring for subsurface geological repositories were discussed and it was shown that the use of muon monitoring for detecting disposal cask movement or unauthorized excavation and removal of rock could be possible.

## Acknowledgments

This work was conducted as a part of the Nuclear Engineering and Science Laboratory Synthesis program at Oak Ridge National Laboratory and was supported by the Office of Defense Nuclear Nonproliferation R&D of the US Department of Energy's National Nuclear Security Administration (DOE/NNSA/NA-22) under contract DE-AC05-00OR22725 in partnership with the NNSA-sponsored University Consortium for Verification Technologies. The authors would like to acknowledge Dr. David Williams, Dr. Stephen Croft, and Dr. Jianwei Hu at Oak Ridge National Laboratory for providing guidance and support to this project, and Andrea Beatty and Kenneth Barker at Oak Ridge National Laboratory for administrative and HPC support.